\documentclass[conference]{IEEEtran}
\IEEEoverridecommandlockouts
\usepackage{cite}
\usepackage{amsmath,amssymb,amsfonts}
\usepackage{graphicx}
\usepackage{textcomp}
\usepackage{xcolor}
\usepackage{hyperref}
\def\BibTeX{{\rm B\kern-.05em{\sc i\kern-.025em b}\kern-.08em
    T\kern-.1667em\lower.7ex\hbox{E}\kern-.125emX}}
    
\usepackage{breqn}
\usepackage{algorithm}
\usepackage{algpseudocode}
\usepackage{pifont}
\usepackage{listings}
\usepackage{float}
\def\BibTeX{{\rm B\kern-.05em{\sc i\kern-.025em b}\kern-.08em
    T\kern-.1667em\lower.7ex\hbox{E}\kern-.125emX}}
    
\begin{document}
\makeatletter
    \newcommand{\linebreakand}{%
      \end{@IEEEauthorhalign}
      \hfill\mbox{}\par
      \mbox{}\hfill\begin{@IEEEauthorhalign}
    }
    \newcommand{\newlineauthors}{%
      \end{@IEEEauthorhalign}\hfill\mbox{}\par
      \mbox{}\hfill\begin{@IEEEauthorhalign}
    }
    \makeatother

\title{COVID-19 and Big Data: Multi-faceted Analysis for Spatio-temporal Understanding of the Pandemic with Social Media Conversations}
\author{
\IEEEauthorblockN{Shayan Fazeli}
\IEEEauthorblockA{
\textit{UCLA}\\
Los Angeles, CA \\
shayan@cs.ucla.edu}
\and
\IEEEauthorblockN{Davina Zamanzadeh}
\IEEEauthorblockA{
\textit{UCLA}\\
Los Angeles, CA \\
davina@cs.ucla.edu}
\and
\IEEEauthorblockN{Anaelia Ovalle}
\IEEEauthorblockA{
\textit{UCLA}\\
Los Angeles, CA \\
anaeliaovalle@g.ucla.edu}
\newlineauthors
\IEEEauthorblockN{Thu Nguyen}
\IEEEauthorblockA{
\textit{UCSF}\\
San Francisco, CA \\
thu.nguyen@ucsf.edu}
\and
\IEEEauthorblockN{Gilbert Gee}
\IEEEauthorblockA{
\textit{UCLA}\\
Los Angeles, CA \\
gilgee@ucla.edu}
\and
\IEEEauthorblockN{Majid Sarrafzadeh}
\IEEEauthorblockA{
\textit{UCLA}\\
Los Angeles, CA \\
majid@cs.ucla.edu}
}
\maketitle
\begin{abstract}
COVID-19 has been devastating the world since the end of 2019 and has continued to play a significant role in major national and worldwide events, and consequently, the news. In its wake, it has left no life unaffected. Having earned the world's attention, social media platforms have served as a vehicle for the global conversation about COVID-19. In particular, many people have used these sites in order to express their feelings, experiences, and observations about the pandemic. We provide a multi-faceted analysis of critical properties exhibited by these conversations on social media regarding the novel coronavirus pandemic. 
We present a framework for analysis, mining, and tracking the critical content and characteristics of social media conversations around the pandemic.
Focusing on Twitter and Reddit, we have gathered a large-scale dataset on COVID-19 social media conversations. Our analyses cover tracking potential reports on virus acquisition, symptoms, conversation topics, and language complexity measures through time and by region across the United States. We also present a BERT-based model for recognizing instances of hateful tweets in COVID-19 conversations, which achieves a lower error-rate than the state-of-the-art performance. Our results provide empirical validation for the effectiveness of our proposed framework and further demonstrate that social media data can be efficiently leveraged to provide public health experts with inexpensive but thorough insight over the course of an outbreak.
\end{abstract}


\section{Introduction}
In the wake of the rapidly spreading pandemic, COVID-19 has stormed to the center stage of the world. ICUs and hospitals have been bombarded with COVID-19 cases in a short amount of time as healthcare providers and researchers seek to further understand and mitigate the effects of COVID-19. 
This endeavor requires collecting as much data as possible surrounding the disease, including significant public health outcomes.
The works on tasks such as predicting or detecting COVID-19 and prognosis for mortality risk using lab-work, or electronic health records, have contributed to an arsenal of tools to help mitigate the effects of the pandemic \cite{vaidFederatedLearningElectronic2020,Wynantsm1328}.
Though electronic health records provide a valuable resource for making inferences regarding the physical effects of the disease, gaining an in-depth insight into how a pandemic at this scale impacts lives is not limited solely to health data from hospitalized patients.

With social media platforms now becoming an inseparable part of daily life for many, a majority of internet users spend considerable time sharing content and their points of view with other members.        
Given the plethora of chatter occurring, our goal is to dig into COVID-19 related conversations taking place across the US and unearth and monitor the key patterns that can shed light on the focus and orientation of public opinion regarding the matters related to the pandemic. These components include the major topics of discussion in the COVID-19 conversations, self-reports of virus acquisition, mentions of explicit and implicit symptoms, as well as generation and spread of hate speech content related to the pandemic and conspiracy theories. We concentrated on sampling and monitoring data from specific geographical regions (US states), which enables the formation of spatio-temporal signals corresponding to each region. As the pandemic impacts are not uniform across all regions (e.g., New York was affected earlier and with a higher severity at the onset of the pandemic), such analyses are consequential for effective tracking of pandemic impacts on each community and ultimately on the broader society.

In this work, we present a comprehensive tracking and analytics framework for monitoring pandemic-related conversations across regions, which addresses all of the key components required for such platforms, as mentioned above.

First, we explored conversation topics that co-occur with COVID-19 as they provide qualitative insight into subjects and/or themes that people are concerned about as they relate to COVID-19.
To this end, we perform topic discovery via Latent Dirichlet Allocation (LDA). 

Second, We monitored self-reported symptoms in order to capture symptoms of those affected by COVID-19 that might be able to or do not choose to get medical care, as well as to capture those that may not directly have contracted COVID-19 but may know someone who did.
In order to capture symptoms, we are tasked with entity discovery, which we effectively tackle by using our designed Recursive Transition Networks (RTNs), such as the one shown in Figure \ref{fig:symptom_graph}.
In addition to monitoring and recognition of such instances within Tweets through time and across different regions, we performed a targeted acquisition of posts in Reddit as another instance of a major social media platform.
These data have helped us present a comprehensive statistical summary on the mentioned symptoms within online COVID-19 conversations.

Additionally, high-level attributes of language complexity are essential indicators for a variety of objectives, such as mental fatigue or long-term health outcomes \cite{snowdon1996linguistic,danner2001positive}.
Therefore, we attempted to monitor these features across the entire data on the pandemic-related conversations.

Lastly, we prepared a state of the art machine learning model as another main contribution which is trained on the task of recognizing hate speech instances in the pandemic tweets.
COVID-19 does not only appear to have different health outcomes for different racial/ethnic groups \cite{szeEthnicityClinicalOutcomes2020} but also social outcomes.
% Given the rise in instances of COVID-19 related hate speech in social media \cite{ziems2020racism}, tracking the severity of such incidents across different regions and through time as different events related to the epidemic occur is a critical matter.
Furthermore, wider impacts such as the emergence of COVID-19 stigma have been causes for concern.
Derogatory terms such as "{\it kung flu}" only started to appear in tweets after the coronavirus outbreak.
Given the surge in hateful comments and evolution of terminology, further monitoring capability of discriminatory language is needed.
It is particularly critical to monitor the spread of hate speech related to the pandemic in social media across regions and be able to cross-reference the fluctuations and patterns in it with various news articles and events.
Leveraging geo-located social media data can facilitate the cross-referencing of COVID-19 conversations with events that are potentially influential. This objective has numerous critical public health applications as the spread of stigma and conspiracy theory can impact the tendency of minority communities to seek medical help when needed.

Considering that social media platforms are a widely used outlet where users express their views on different matters and events, event-centered analysis on the use of language is incredibly valuable in assessing health and broader impacts of the pandemic. 

\subsection{Related Work}
Provided its ubiquity, a multitude of research has focused on extracting information from social media data to understand the links between online behavior and health outcomes. For instance, work has been done regarding the manifestation and impacts of "social media addiction" on people's lives and their mental health \cite{nakaya2015internet,hawi2017relations,liu2018social,leong2019hybrid,esgi2016development}. Previous literature has also pointed to how social media can be an invaluable resource for health and education while also noting that care must be taken in its respective analyses. \cite{whyte2017social,duymus2017internet}.

Social media data has allowed for numerous effective data mining and knowledge extraction methodologies, such as sentiment analysis and recognition of hate speech, cyber-bullying, and misinformation \cite{florio2020time,mondal2017measurement,malmasi2017detecting,makrynioti2017palopro,ma2019automatic,bode2015related,wu2016mining,allcott2019trends,trottier2012key}.
Now, with COVID-19, several works have incorporated social media data to better understand the disease.
For example, previous work tackles tracking topics of conversation that coincide with COVID-19 chatter, both on social media \cite{ordun2020exploratory} and in academics \cite{algaAnalysisScientificPublications2020}.

Additionally, detecting and tracking symptom reports in social media is an ongoing research area \cite{guntuku2020tracking,al2020text,sarker2020self,marshall2016symptom}.
While focusing on more traditional approaches such as analyzing word frequencies or binary classification, the aforementioned works illustrate the effectiveness of using social media data for symptom discovery.
Mental health and social media consumption has also been explored during COVID-19 \cite{gao2020mental}.
The CORD-19 challenge has also focused on the automatic extraction of entities and information from articles on the COVID-19 pandemic and other pandemics with similar characteristics \cite{wang2020cord}.
Still, further studies are needed with data modalities that are qualitatively diverse and real-time in order to facilitate action by public health experts \cite{german2001updated}.
Moreover, analyses on symptoms and greater public health impacts through time and place are a necessary component of monitoring platforms as experts observe drastic differences in infection rate and government intervention across various regions \cite{dave2021shelter}.

Other datasets surrounding the pandemic on Twitter exist.
Some datasets collect data about COVID-19 in general \cite{lermandata2020}, while others have more specific focuses.
For example, one dataset is a collection of Tweets containing the words ``China'' and ``coronavirus'' \cite{StigmatizationFan2020}.
%% TODO add citations for other datasets?
While there exists other Twitter hate speech datasets that revolve around COVID-19, they might focus on other languages such as Arabic \cite{arabichatecovid}.
By contrast, our dataset focuses on geo-tagged English tweets from the United States that relate to COVID-19.
This implies that our dataset is a subset of pandemic-related tweets that originate from authors who have enabled geo-tagging.

To the best of our knowledge, no other work in this area has focused on a cohesive integration of the spatio-temporal analysis of social media conversations, extraction and tracking of important entities, and the assessment of sentiment and various linguistic capacities.
Our paper expands on existing literature and addresses the aforementioned research gaps to further understand the COVID-19 pandemic and its impacts on society.
To do this, we gathered and analyzed large-scale COVID-19 data from popular social media platforms.
We extracted likely mentions of COVID-19 infections and symptom entities from the raw and unstructured text and shed light on fluctuations of symptoms in conversation trajectories through time, data domains, and events.

In particular, our contributions are as follows:

\begin{itemize}
    \item {
        We created COVISION \footnote{
            Our code is available at \href{https://github.com/shayanfazeli/covid_and_bigdata}{https://github.com/shayanfazeli/covid\_and\_bigdata}
        }, a large-scale dataset containing over $3.5$ million geo-tagged tweets related to COVID-19 across nine months. We additionally collected over $20,000$ Reddit posts on COVID-19 experiences and symptoms to create a diverse source of information on self-reports of symptoms and physical and mental health.}
    \item{
        We utilized our framework for an in-depth analysis of individual posts from the four key aspects of:
        \begin{enumerate}
            \item Self-reported symptoms and virus acquisition
            \item Topic of Conversation related to the COVID-19 (e.g., social distancing)
            \item Language characteristics
            \item Hate and Stigma-related content
        \end{enumerate}
    }
        \item{
        We present a machine learning model based on neural networks achieving the state of the art results on the task of recognizing hate speech in COVID-19 data with a significant increase over the baseline performance.
    }
    \item{
        We created a platform able to form and track aggregate spatio-temporal signals relevant to the corresponding social media activity and demonstrated the presence of critical information in the signals by visualizing the change points cross-referenced with major pandemic-related events in the news.
    }
    \item{ 
        We used a wide set of quantitative and qualitative methods and empirically validated the efficacy of our proposed platform for performing the multi-faceted analysis of pandemic conversation in social media.
    }
\end{itemize}

\begin{table}[]
    \centering
    % captions should be above the table
    \caption{COVISION Statistics}
    \begin{tabular}{|c|c|}
    \hline
        Number of Tweets &  $3,506,405$\\ \hline
        Number of Authors & $750,315$ \\\hline
        Earliest Observed Date & 2020-01-01 \\\hline
        Latest Observed Date & 2020-10-01 \\\hline
    \end{tabular}
    
    \label{tab:covision_stats}
\end{table}

\begin{figure}
    \centering
    \includegraphics[width=\linewidth]{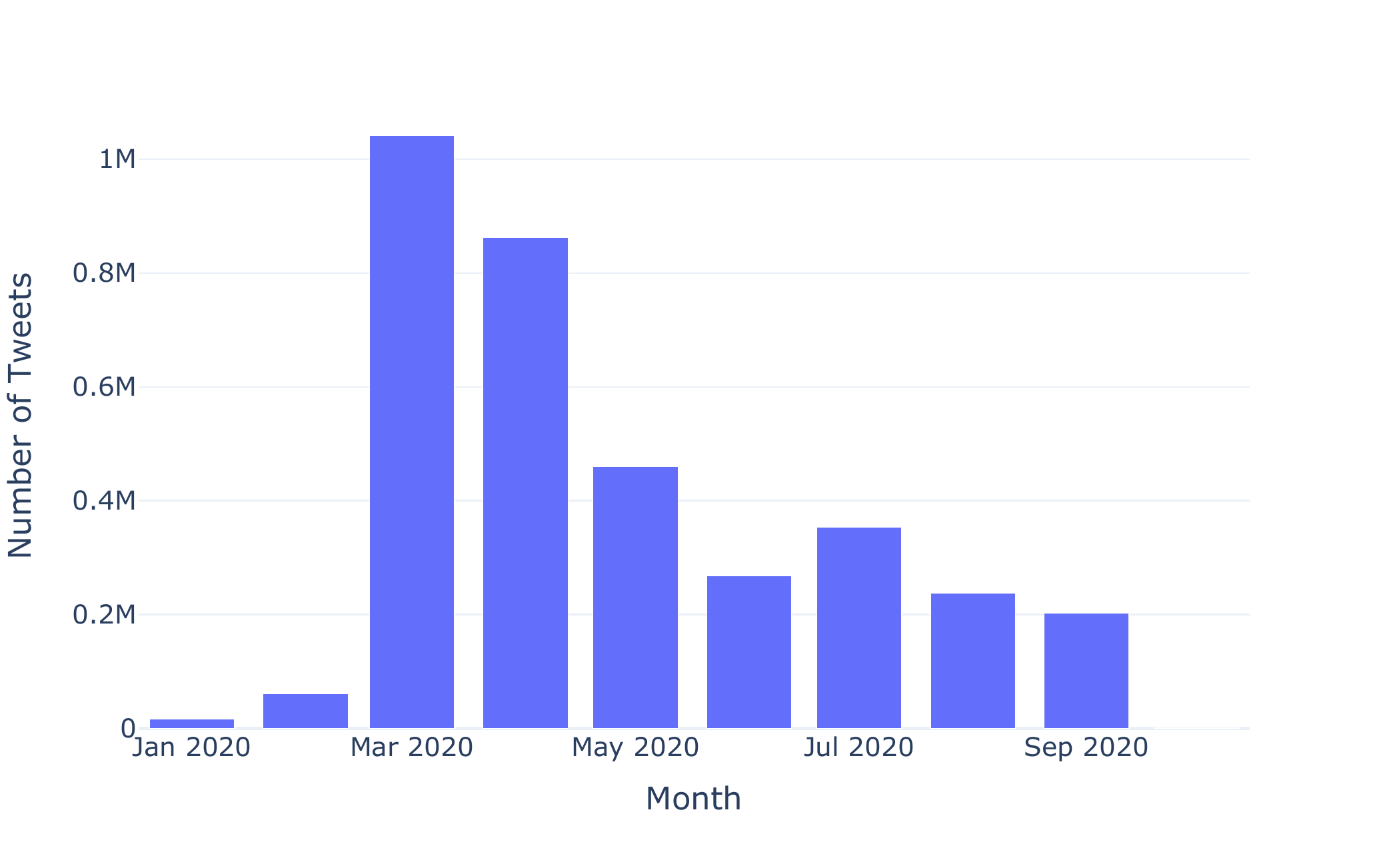}
    \caption{Distribution of COVID-19 tweets per month across the US in the COVISION dataset}
    \label{fig:coviddistribution}
\end{figure}

\section{Methods}

\subsection{Data Collection}
The first step towards building an effective monitoring platform is gathering a large-scale corpus on COVID-19 related data from social media platforms.
While there are several social media datasets presented in the literature, the main focus so far has been on including a large number of tweets. 
In this work, our data acquisition criteria were focused on geo-tagged tweets to enable the formation and analysis of spatio-temporal signals in addition to in-depth probing of individual posts.
Given that geo-tagged tweets comprise a considerably small portion of all the posted tweets, we gathered our data in a long-term span of $9$ months, covering the onset of the pandemic and the first waves. These choices resulted in a corpus of $3.5$ million tweets captured from the publicly available tweets in the date range of January 1st, 2020, until October 1st, 2020. In summary, each tweet in our corpus:
\begin{itemize}
    \item Contains a reference to the COVID-19 pandemic (e.g. "covid19", "quarantine", "chinesevirus", Please refer to the appendix for the full list of our search terms.)
    \item Is posted in the United States.
    \item Contains location information (coordinates or other place attributes) which permits the identification of the US state in which it was posted.
\end{itemize}

Figure \ref{fig:coviddistribution} shows the distribution of tweets per month, and as expected, the peak appears in March and April of 2020 as the pandemic problem was starting to worsen across the US.
Many states in the US had not issued stay-at-home order until early to mid-March, which is when COVID-19 became a jarring centerpiece of conversation for many people.
As everyone adjusted to life with COVID-19, we expect, and ultimately observe, that conversation surrounding the pandemic observed a reduction in the number of geo-tagged tweets in the coming months.

The number of tweets is not evenly distributed across the months, which is a reasonable pattern. The number of tweets is an informative signal given that limiting our data to geo-tagged corpus leads us to gather a representative subsample of the overall distribution of conversations. Therefore, we do not observe the saturation in tweet counts due to API rate-limits.

%% reddit
As suggested in the literature, there are significant differences in people's approach towards social media based on the type of platform \cite{kennedy2020constructing}. 
To add to the diversity of our data reflecting on the self-reports, we additionally focused on the Reddit platform to obtain more data regarding people's experiences with COVID-19.
Given the level of anonymity that Reddit provides, one could argue that users find it easier to share more detailed descriptions of their experiences on this platform.
In order to limit our search scope to COVID-19 experiences, we focused on the "{\it AskReddit}" subreddit.
"{\it AskReddit}" is a popular sub-community on Reddit dedicated to threads where a poster poses a question, often personal, and users participate and answer the question by sharing their experiences.
We gathered $23,770$ responses to questions about the COVID-19 pandemic experience and symptoms experienced by users or people close to them.
It is worth mentioning that we chose these threads based on the attention they received from the broader user-base.
Also, Reddit primarily functions as a forum, therefore, posts are not geo-tagged and conversations are focused on a singular subject.
While this format does not allow us to utilize this data in improving the spatio-temporal conversation trajectories, it provides a great source of information for refining our findings on symptoms and self-reports of experience with COVID-19.

% Do note that given the format of the platform as a forum, posts are not geo-tagged and often the conversation is clustered around a the time point (time of posting) until the thread loses popularity within a few days and becomes stale (no longer experiences activity).
% Therefore, even though the data from Reddit are valuable in ameliorating our dataset and analysis, we can only apply it towards tasks such as symptom and topic extraction.

\subsection{Post Probing}
\subsubsection{Self-reports}
One of our primary tasks was extracting mentions of COVID-19 acquisition and symptom entities from the raw and unstructured text.
We designed topic-specific Recursive Transition Networks (RTN) to look for corresponding manifestations in the text.
RTNs are graph structures describing different rules of a context-free grammar. 
Our grammar-based designs help in a targeted search for instances of terms that are likely associated with reports of virus acquisition or experienced symptoms.
Figure \ref{fig:symptom_graph} demonstrates our RTN instance used for extracting symptoms.

As mentioned before, the social media data in COVISION are acquired focusing on the conversation around the subject of pandemic. 
This relationship means the presence of COVID-19 as the main underlying context in interpreting these data is already established.
Therefore, a grammar-based search is a viable and efficient approach to exploring tweets and discovering patterns that indicate a high likelihood of an important entity being mentioned.

\begin{figure}[h!]
    \centering
    \includegraphics[width=0.49\textwidth]{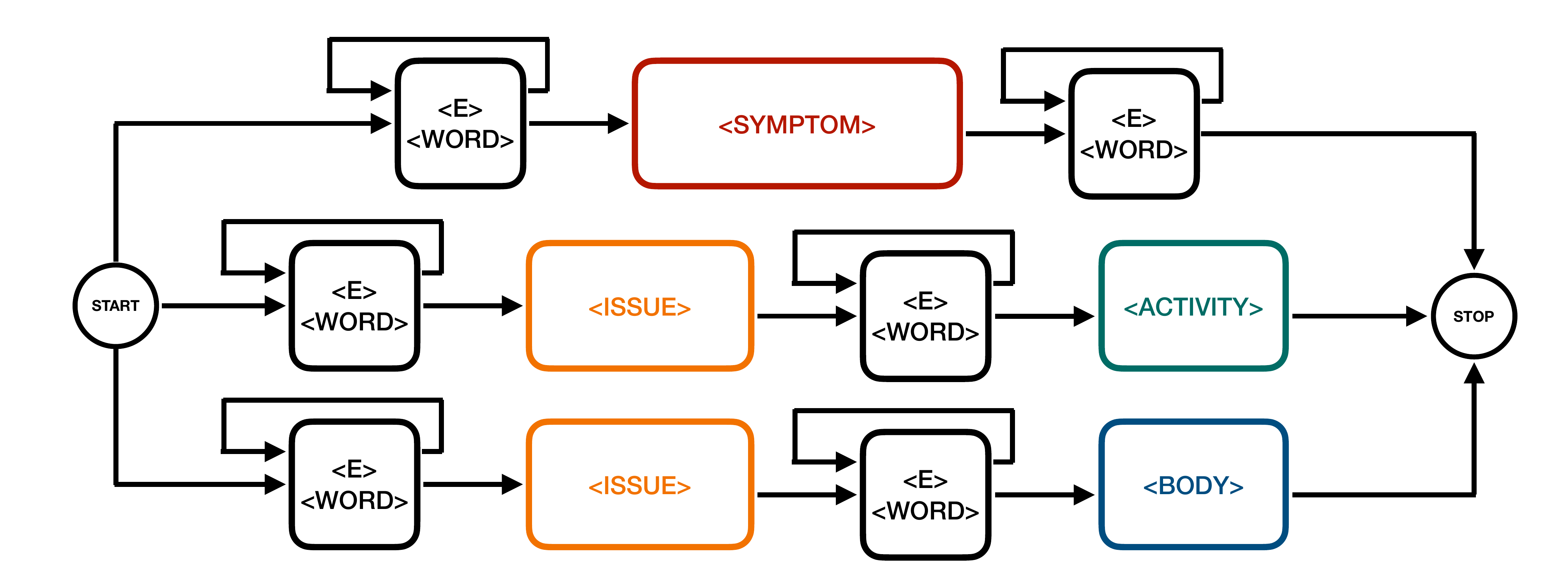}
    \caption{Our Recursive Transition Network for Symptom Search Grammar}
    \label{fig:symptom_graph}
\end{figure}

%% table for symptoms
{\small
\begin{table}[h]
    \centering
    \caption{The grammar-based tags and examples.}
    \begin{tabular}{|c|c|c|}
    \hline
        \textbf{Tag} & \textbf{Description} & \textbf{Example} \\ \hline
        $\langle$ symptom $\rangle$ & Medical Symptom & I have a \textit{headache} \\ \hline
        $\langle$ issue $\rangle$ & Issues in Activity or Body  & I have {\it pain in my neck} \\ \hline
        $\langle$ activity $\rangle$ & Physical Activity & For me it is {\it hard to  breathe} \\ \hline
        $\langle$ body $\rangle$ & Body Parts & I have {\it tremor in my  hand} \\ \hline
    \end{tabular}
    \label{tab:grammar_examples}
\end{table}}

We designed RTN instances for four main categories of COVID-related entity discovery.
The main RTNs shown in Figure \ref{fig:symptom_graph} were used to tag tweets for potential mentions of physical and mental symptoms.
Table \ref{tab:grammar_examples} contains further descriptions of each tag we defined in the RTNs and what they might look like in free text.
A separate RTN was designed in the same fashion to look for possible reports of COVID-19 infection.

The outputs of this methodology provide information on the nature and quantity of entities reported in tweets that are potentially associated with COVID-19. This data-driven approach helps better understand the symptoms and characteristics exhibited by those who contracted COVID-19 and sheds light on the less noticeable aftermath of the disease and lengthy quarantines and shutdowns, such as the increase in the manifestation of depression and quarantine fatigue.

\subsubsection{Discussed Topics}
Given that all of the tweets in the COVISION dataset are related to the COVID-19 pandemic as the primary data acquisition criteria, we focused our efforts on discovering the topics of discussion in these tweets.
We explored a range of numbers for the topics present in this dataset, and used Latent Dirichlet Allocation (LDA) to find, associate, and interpret these topics.
LDA is a popular technique for topic modeling and models documents as a combination of $k$ latent topics and topics as a combination of words (following a unigram language model) \cite{bleiLatentDirichletAllocation2003}.
LDA treats text as an unordered "bag of words" and assumes that the document-to-topic vector of probabilities and topic-to-word vector of probabilities are both sampled from the Dirichlet distribution.

One of the most common metrics used to evaluate the performance of LDA is perplexity.
Perplexity, defined in Equation \ref{eq:perplx}, is inversely related to the likelihood of observing the held-out data.

\begin{equation}\label{eq:perplx}
    \text{perplexity}=\exp(- \cfrac{\sum_{i=1}^T \ln{\Pr(\mathbf{w^{(i)}}  | \Theta)}}{\sum_{i=1}^T N_i})
\end{equation}

In the above formulation, the held-out data is composed of $T$ documents. $\mathbf{w}^{(i)}$ is the sequence of words for the $i$th document, and $\Theta$ is the set of model parameters. Also, $N_i$ is the number of keywords in the $i$th document.
In this case, the goal of the model is to maximize the probability of observing held-out data, for which a common practice is to minimize perplexity.
If the learned topics capture the true latent topics well, the model will generalize and assign a high likelihood of observing the test data documents, resulting in a lower perplexity score.

To find the best number of topics with LDA, we monitored the perplexity trend across a varying number of topics, performed additional qualitative assessments, and chose the number of topics to consider and interpret accordingly.

\subsubsection{Readability}
Measures related to the complexity of post language have previously been shown in the literature to be correlated with a person's physical and mental well-being in the long run.
We employ several well-known metrics in assessing every tweet in terms of complexity and readability, and augment the tweet datasets using these metrics \cite{mc1969smog,smith1967automated,dale1948formula,kincaid1975derivation,coleman1975computer,newsairforce,o1966gobbledygook,10.2307/998915}.
These scores are not sensitive to text order and they use a count-based approach to determining the score, which often requires long text data.
Therefore we first aggregated by time and region to reflect on the overall readability of the conversation, which can be thought of as a large text to which many authors have contributed.
Do note that the language complexity measures are agnostic to the number of authors for a piece of text.
Please refer to section \ref{apx:langcomplexity} of the appendix for more details on these algorithms.

\subsubsection{Sentiment}
The analysis of the sentiment in public posts associated with COVID-19-related hate and stigmatization is another important component of our framework.
Given the rise in the number of hate-related posts across social media platforms since the onset of the pandemic, an effective monitoring mechanism is required to be able to detect which posts are hate speech-related, including both hateful and counter-hate tweets.
A tweet that is labeled as counter-hate denounces a hateful opinion.
This helps to shed light on the patterns of hate speech generation and how it is spread, which is especially helpful in a pandemic as conspiracy theories and stigma associated with pandemic-related hate speech can lead to disparities against minority communities \cite{budhwani2020creating}.
Using our framework of analysis allows the experts to optimize the required resource allocation to alleviate such negative impacts of a pandemic. 

Recently, CLAWS, a large dataset on COVID-19 hate speech on Twitter, was released \cite{ziems2020racism}.
This dataset includes more than $3$ million tweet ids along with the corresponding hard labels of {\it hate}, {\it neutral}, or {\it counterhate}.
We focused on this dataset to train our hate speech modules and evaluate the performance of our platform in recognizing hate speech content on Twitter in the midst of conversation about COVID-19.
This is due to the fact that there is a domain shift in the hateful content related to the COVID-19 and what was published in the hate speech datasets before the pandemic.
After crawling the tweet ids, we obtained $2.5$ million tweets with labels, a number that is still considerably larger than the cardinality of many of the older hate speech datasets (which contain tens of thousands of tweets).

We propose a deep inference pipeline to process the tweets and associate a label with them according to the task defined in \cite{ziems2020racism}, marking them as hateful, counter-hate, or neutral.
Our neural network architecture is composed of the standard BERT transformer architecture initialized with the pre-trained weights \cite{devlin2018bert}, along with a shallow classification layer based on fully connected neural networks on top of the average pooled BERT representations.
Given that the CLAWS dataset is relatively limited in size, this allows us to utilize inductive transfer learning and obtain more efficient semantic representations in terms of generalization and performance.
We then employ gradients to train and fine-tune the network, including the core language model, on the CLAWS task.

After training and evaluation of our hate speech recognition module, we followed the protocol in \cite{ziems2020racism} and defined the CLAWS task as labeling a tweet with one of the three labels: $\{\text{neutral},\text{hate},\text{counter-hate}\}$.
This was then used as part of our platform to employ on our gathered data, to provide another critical signal from the information each post contains, and to also help with forming the corresponding aggregate spatio-temporal trajectories.

Due to the fact that we trained our model using the CLAWS dataset, we were careful regarding the change in domain distribution while applying the model to our gathered data.
We qualitatively assessed the results on our data and did not observed an explicit domain shift.
This is an expected observation as most of the new instances of hate speech (e.g., derogatory terms such as {\it kung flu}) were introduced in the onset of the pandemic and remained relatively the same throughout the coming months.
In addition, the CLAWS dataset includes a large number of tweets from the United States and is therefore expected to cover enough material for our model to effectively learn the task of hate speech labeling\footnote{It is noteworthy that our data and CLAWS dataset are almost entirely independent, as less than $0.2 \%$ of our tweets can be found in the subset of the CLAWS dataset used in our work.}.

\subsection{Spatio-temporal Aggregation}
Similar to the case of monitoring pandemic outcomes within subregions (e.g., time-series of the number of cases per state), aggregating information and forming temporal signals associated with each region helps better illustrate the regional variations in the characteristics of the conversation surrounding the topics related to the pandemic. In our platform, aggregated information on the number of self-reported entities, hate/stigma-related sentiment, and language complexity were used to form these temporal signals. We also considered the major events related to the pandemic from news outlets and observed that cross-referencing such events can illustrate potential relationships and help experts form and test hypotheses regarding the pandemic conversations.

\begin{figure*}[h]
    \centering
    \includegraphics[width=0.95\textwidth]{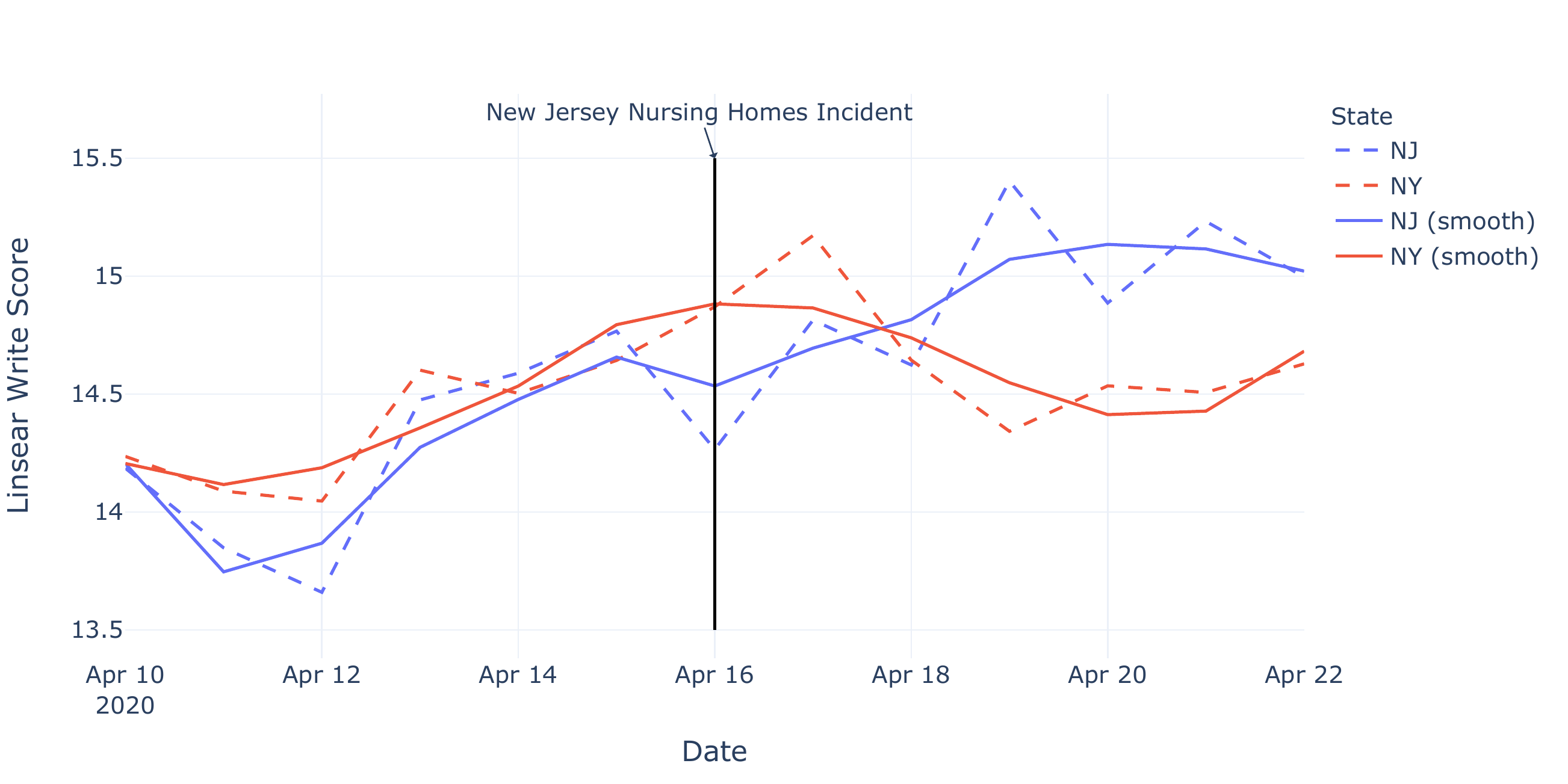}
    \caption{Daily Average of Linsear Write Score for New York and New Jersey showing a decrease in overall readability}
    \label{fig:langcomplexity_nynj}
\end{figure*}

\begin{figure*}[h]
    \centering
    \includegraphics[width=0.95\textwidth]{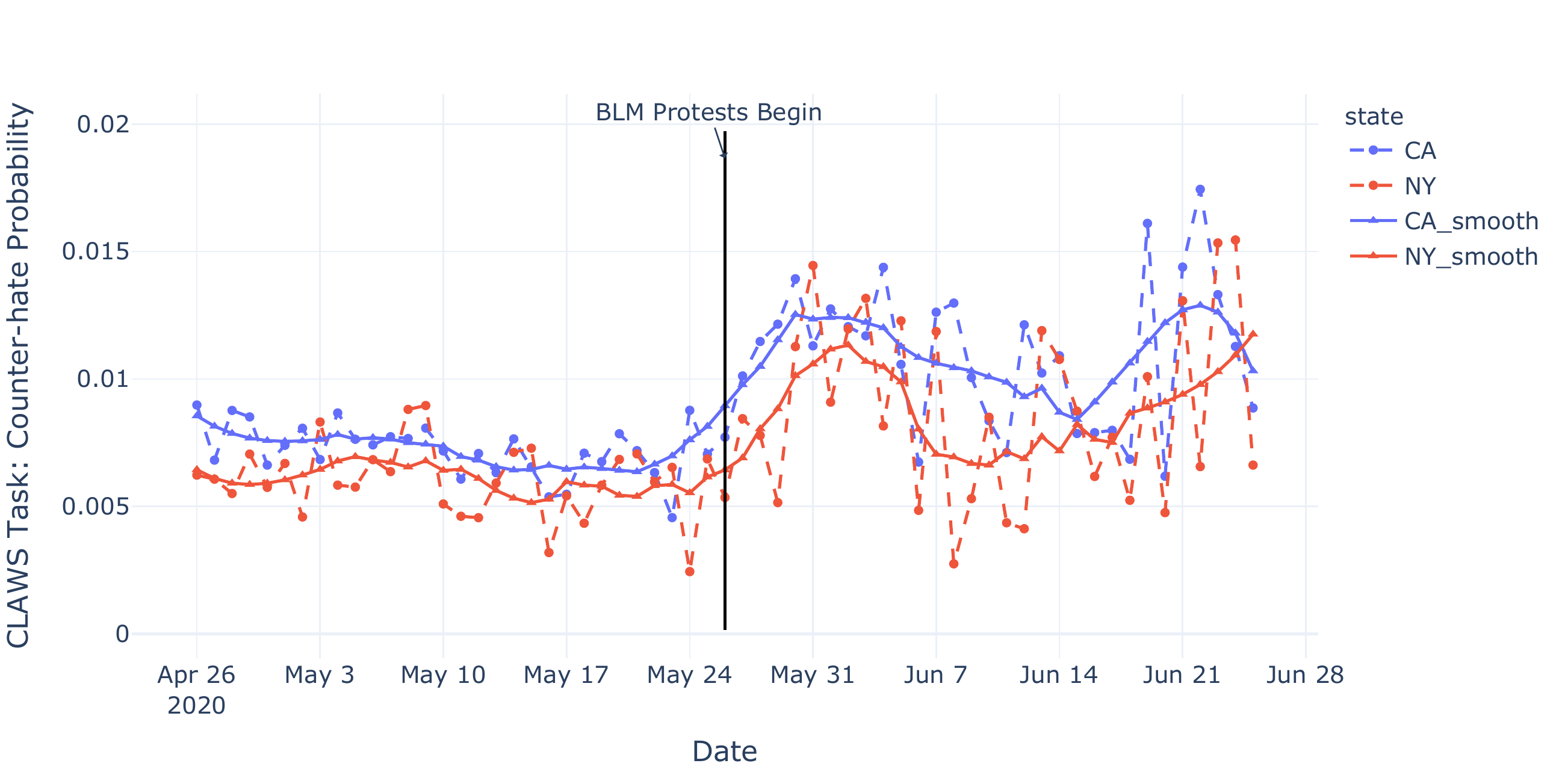}
    \caption{Daily Average of CLAWS Task: Counter-hate Probability around event: BLM Protests Begin - Half Window: 30 days}
    \label{fig:median_counterhate}
\end{figure*}

\section{Results and Discussion}

\subsection{Post Probing}
\subsubsection{Self-reports}
We designed our RTNs for a grammar-based searching scheme in UniTex GramLab version 3.2 \cite{unitexgramlab}.
We then tagged the tweets by employing the designed RTNs so as to extract the reported symptoms and infections from the Twitter and Reddit data.
Thorough examples are shown in Figure \ref{fig:tagged_tweet_orig} and Figure \ref{fig:tagged_reddit0}, where it is shown that even an extensive list of potential symptom entities within a single post can be extracted with ease.

Table \ref{tab:symptoms_twitter_and_reddit} shows the top $10$ most frequently mentioned symptoms in the COVID-19 conversation across Twitter and Reddit posts in our dataset, respectively.

% \begin{table}[]
%     \centering
%     \caption{Most frequent potential symptoms mentioned in Twitter posts of our dataset}
%     \label{tab:symptoms_twitter}
%     \begin{tabular}[|ll|]
%         \hline
%         a & b \\
%         \hline
%     \end{tabular}
% \end{table}

\begin{table}[]
    \centering
\caption{Most frequent potential symptoms mentioned in posts in our dataset}
\label{tab:symptoms_twitter_and_reddit}
\begin{tabular}{l|l|l|l}
\textbf{Term (Twitter)} & \textbf{Freq.} & \textbf{Term (Reddit)} & \textbf{Freq.} \\ \hline
tired	& $24.29 \%$ & fever & $27.67 \%$  \\
cough	& $15.24 \%$ &  cough & $23.05 \%$ \\
depression	& $14.59 \%$ & coughing & $8.51 \%$ \\
fever	& $13.27 \%$ & fatigue	& $6.60 \%$ \\
coughing	& $7.47 \%$ & aches & $6.33 \%$ \\
weak	& $7.34 \%$ & shortness of breath	& $4.38  \%$ \\
fatigue	& $2.16 \%$ & tired & $4.30  \%$ \\
coughed	& $1.76 \%$ & chills & $2.83 \%$ \\
aches	& $1.43 \%$ & weak & $2.69 \%$ \\
shortness of breath	& $1.42 \%$ & diarrhea & $2.63 \%$ \\
\end{tabular}
\end{table}

% \begin{table}[]
%     \centering
% \caption{Most frequent potential symptoms mentioned in Twitter posts of our dataset}
% \label{tab:symptoms_twitter}
% \begin{tabular}{l|l}
% \textbf{Term} & \textbf{Freq.} \\ \hline
% tired	& $24.29 \%$ \\
% cough	& $15.24 \%$ \\
% depression	& $14.59 \%$ \\
% fever	& $13.27 \%$ \\
% coughing	& $7.47 \%$ \\
% weak	& $7.34 \%$ \\
% fatigue	& $2.16 \%$ \\
% coughed	& $1.76 \%$ \\
% aches	& $1.43 \%$ \\
% shortness of breath	& $1.42 \%$ \\
% \end{tabular}
% \end{table}

% \begin{table}[]
%     \centering
% \caption{Most frequent potential symptoms mentioned in Reddit posts of our dataset}
% \label{tab:symptoms_reddit}
% \begin{tabular}{l|l}
% \textbf{Term} & \textbf{Freq.} \\ \hline
% fever & $27.67 \%$ \\
% cough & $23.05 \%$ \\
% coughing & $8.51 \%$ \\
% fatigue	& $6.60 \%$ \\
% aches & $6.33 \%$ \\
% shortness of breath	& $4.38  \%$ \\
% tired & $4.30  \%$ \\
% chills & $2.83 \%$ \\
% weak & $2.69 \%$ \\
% diarrhea & $2.63 \%$ \\
% loss of taste & $2.08 \%$ \\
% \end{tabular}
% \end{table}

\begin{figure}[H]
    \centering
    \begin{lstlisting}[basicstyle=\ttfamily \footnotesize]
>> tweet (raw): I cant sleep early again, this covid 19 has got me all messed up! Fever, chills, headache, tired, cough, shortness of breath, no sense of smell, nausea, diarrhea, my eye socket hurts. No wonder some people just give up or the body just gives up.

>> tweet (tagged): cant sleep early again this covid has got me all messed up <symptom>fever<symptom> <symptom>chills<symptom> headache <symptom>tired<symptom> <symptom>cough<symptom> <symptom>shortness of breath<symptom> no sense of smell <symptom>nausea<symptom> <symptom>diarrhea<symptom> my eye socket hurts no wonder some people just give up or the body just gives up.

<symptom>: fever::chills::tired::cough::shortness of breath::nausea::diarrhea
<covid_report>: none
<impact_body>: none
<impact_activity>: none
    \end{lstlisting}
    \caption{Example tagged tweet using RTN}
    \label{fig:tagged_tweet_orig}
\end{figure}

\begin{figure}[H]
    \centering
    \begin{lstlisting}[basicstyle=\ttfamily \footnotesize]
>> reddit comment (tagged): My fiancee is an ICU nurse, so of course we both <covid_report>got it<covid_report> (and our toddler <covid_report>got it<covid_report>).It sucked. Mostly for me; toddler got through it with barely any effects, my fiancee got through it with being mildly ill and <symptom>tired<symptom> for a week or so...I thought I had a pretty mild case until suddenly I got winded by walking ~10 meters. I was just barely not sick enough to get admitted to the hospital for it. Apart <covid_report>from that<covid_report>, the worst thing for me was the <symptom>fatigue<symptom>.I had symptoms for around 2 weeks, and the middle-to-last part I was just so damn <symptom>tired<symptom> all the time. A few days, I slept around 15-16 hours out of the day. We lost our sense of taste and smell, though they are *mostly* back; taste is completely back, but sometimes it is <impact_activity>hard to smell<impact_activity> the scent of things.

<symptom>: tired::fatigue::tired
<covid_report>: got it::got it::from that
<impact_body>: none
<impact_activity>: hard to smell
    \end{lstlisting}
    \caption{Example tagged Reddit post using RTN}
    \label{fig:tagged_reddit0}
\end{figure}

\begin{figure}
\centering
\includegraphics[width=0.45\textwidth]{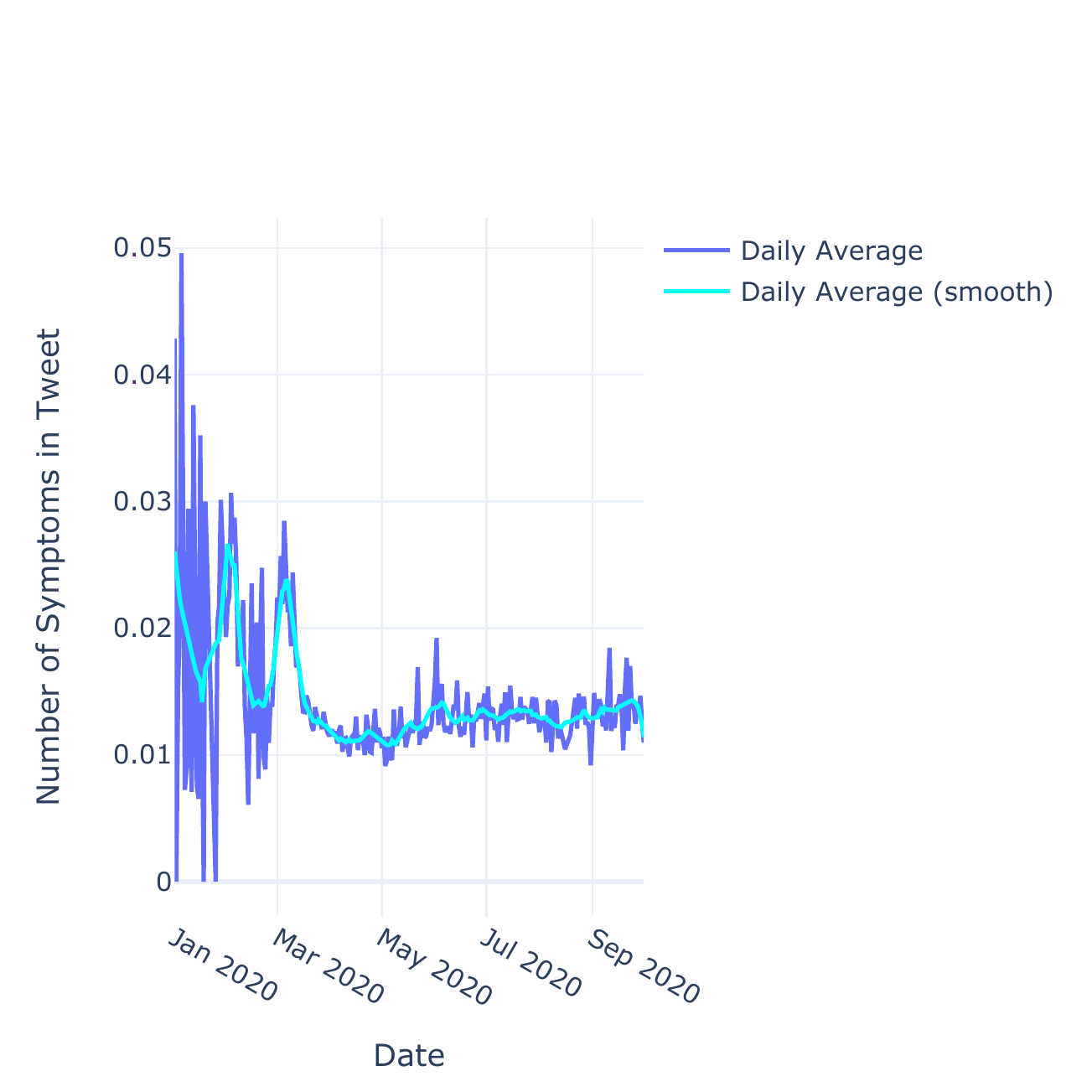}
\caption{Average number of symptom entities in tweets through time}
\label{fig:symptom_count}
\end{figure}

\begin{figure*}
    \centering
    \includegraphics[width=0.95\textwidth]{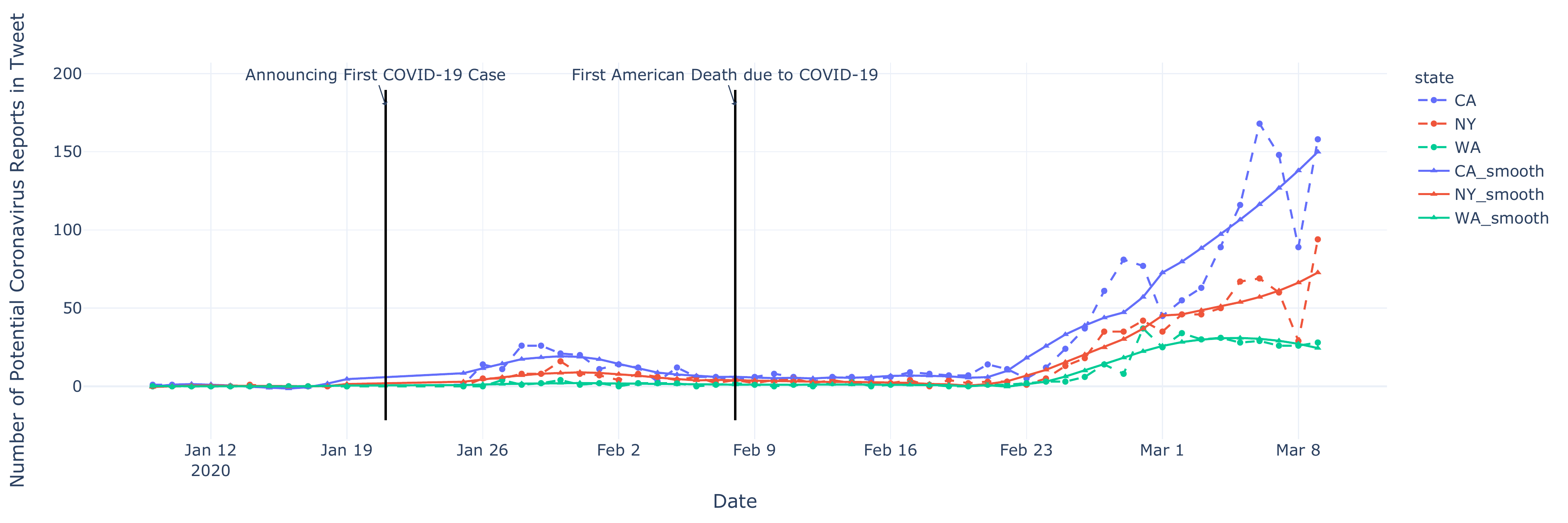}
    \caption{Potential COVID-19 acquisition reports in major states at the onset of the pandemic}
    \label{fig:covidreportrise}
\end{figure*}

\subsubsection{Discussed Topics}
We trained our model for assigning topics to each post by employing a full-batch LDA algorithm, considering unigrams, bigrams, trigrams and omitting words that appear fewer than $50$ times in the dataset.
We also lemmatize the words as an essential preprocessing step to help improve topic discovery.
The results of the quantitative measurements of performance as perplexity on the hold-out set, in addition to our qualitative assessments, indicate that the model with around $20$ topics is best to fit our Twitter corpus.
The term probabilities per topic are displayed in Figure \ref{fig:twitter_lda_topics_barplots} for our $22$ topic instance.
The qualitative evaluation of the topics and posts led us to consider the topic labels mentioned in Figure \ref{fig:twitter_lda_topics_barplots} as the prominent conversation centroids for the tweets in our dataset.

\begin{figure}
    \centering
    \includegraphics[width=0.49\textwidth]{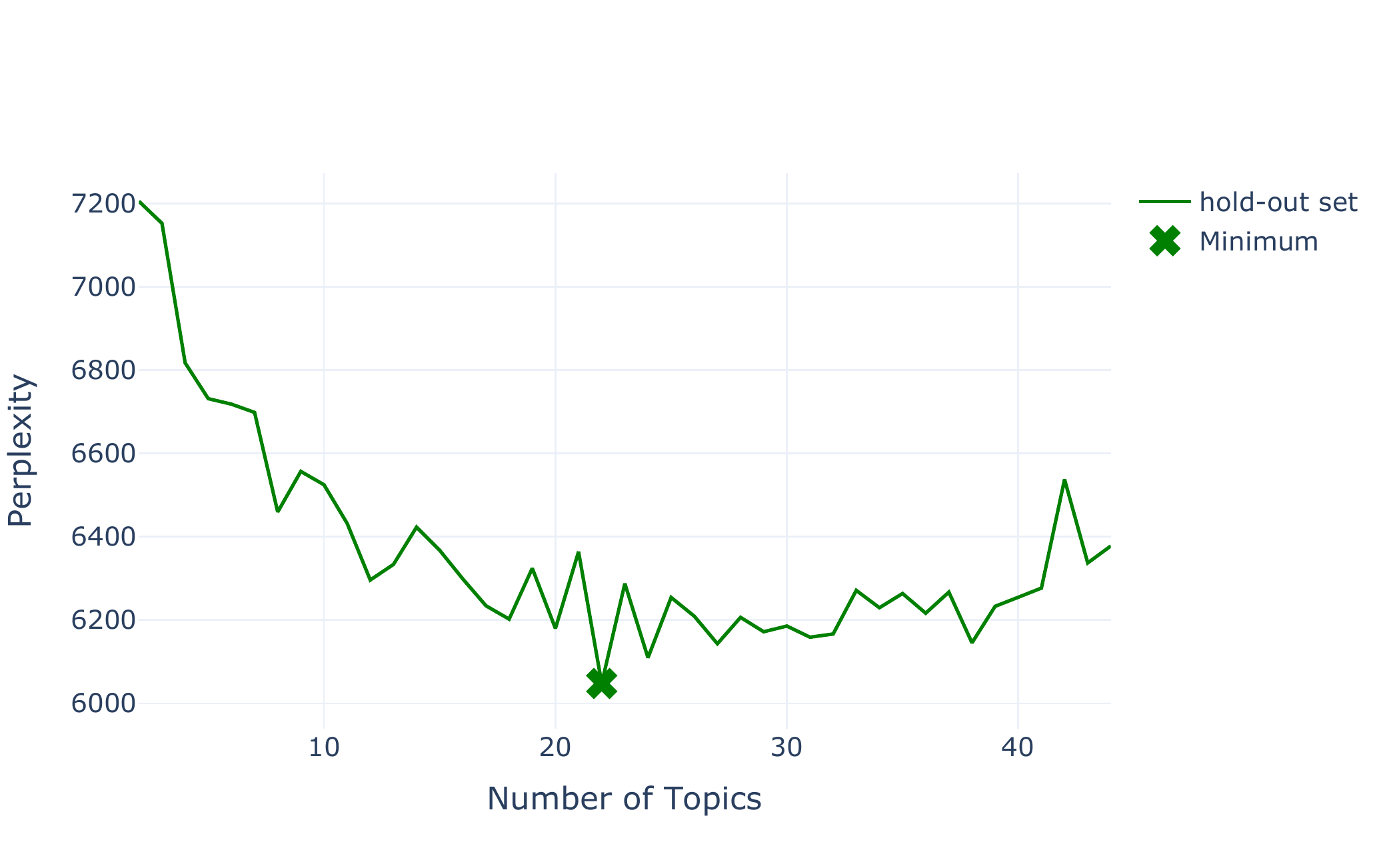}
    \caption{Perplexity vs Number of Topics in our Twitter corpus - Full Batch LDA Topic Modeling}
    \label{fig:perplexities}
\end{figure}

\begin{figure*}
    \centering
    \includegraphics[width=0.79\textwidth]{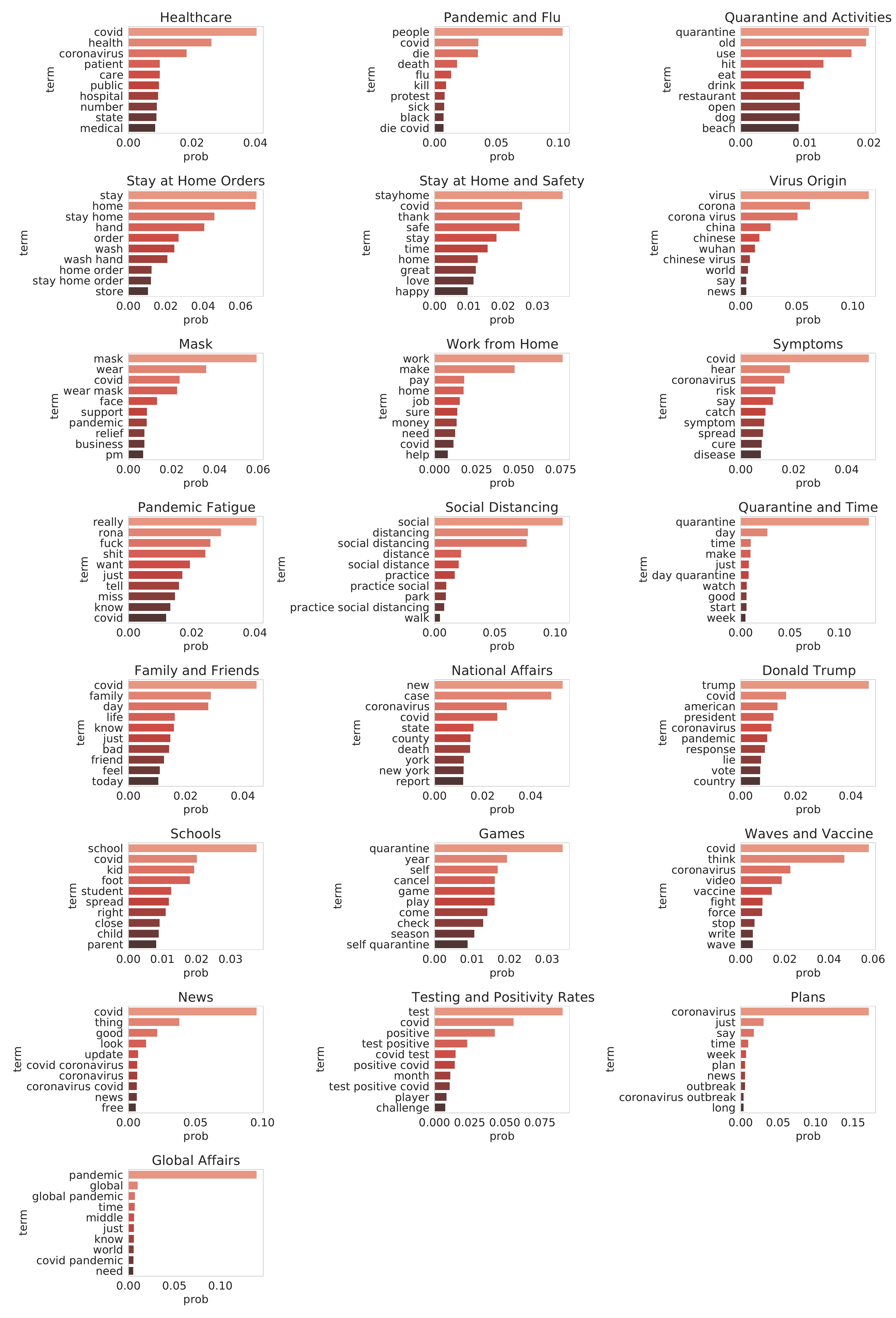}
    \caption{An overview of the main topics of the COVID-19 conversation around the US during the first 9 months of the year 2020.}
    \label{fig:twitter_lda_topics_barplots}
\end{figure*}

We performed the topic modeling analysis with LDA on our Reddit corpus as well. The perplexity values computed on a hold-out set are shown in Table \ref{tab:perp_reddit}. We qualitatively assessed the discovered topics, and the version with $3$ topics appeared to be the best fit for this data.
Table \ref{tab:topics_reddit} outlines some topics discovered and their respective terms.

\begin{table}[]
    \centering
\caption{Topics from LDA Topic Modeling using Reddit Data from COVISION}
\label{tab:topics_reddit}
\begin{tabular}{|c|c|}
\hline
\textbf{Topic Name} & \textbf{Terms} \\ \hline
Acquisition         & symptom, cough, fever, smell, taste  \\ \hline
Support             & family, home, friend         \\ \hline
Cost and Response   & healthcare, hospital, insurance, cost, bill         \\ \hline
\end{tabular}
\end{table}

\begin{table}[]
    \centering
\caption{Perplexities on a hold-out validation set for our Reddit corpus}
\label{tab:perp_reddit}
\begin{tabular}{|c|c|}
\hline
\textbf{Number of Topics} & \textbf{Perplexity} \\ \hline
2                         & 1883.65             \\ \hline
3                         & 646.680             \\ \hline
4                         & 2021.64             \\ \hline
5                         & 2231.07             \\ \hline
\end{tabular}
\end{table}

It is worth noting that the hold-out set for both Twitter and Reddit corpora in COVISION was done by shuffling and splitting author ids, ensuring that each set is composed of different users creating the posts.

\subsubsection{Readability}
% table for readability
Following the literature regarding monitoring language complexity as an indicator for the long-term health-related issues, we assessed the COVISION data on several language complexity measures and a text readability measure to understand temporal patterns that crop up.

As visible in Table \ref{tab:langcomp}, there is a general increasing pattern in the language complexity indicative of a slight decrease in overall tweet readability at the onset of the COVID-19 pandemic in the US.
While not a significant overall increase, this difference can be associated with concepts such as public anxiety or quarantine fatigue.
The upward trajectories of Linsear Write in Figure \ref{fig:langcomplexity_nynj} is another example showing a noticeable increasing pattern in the language complexity in New York. Such fluctuations are important to keep track of as they could be potentially correlated with health outcomes.

\begin{table*}[h!]
    \centering
    \caption{Measures of Readability - Monthly Average Scores across the United States}
\begin{tabular}{|l|l|l|l|l|l|l|l|l|}
\hline
\textbf{Month}     & \textbf{Coleman-Liau Index} & \textbf{Flesch-Kincaid} & \textbf{Dale-Chall} & \textbf{Gunning-Fog} & \textbf{SMOG} & \textbf{SPACHE} & \textbf{Linsear-Write} & \textbf{ARI} \\ \hline
January   & 10.95              & 10.83                & 9.11             & 13.74             & 13.18      & 6.76         & 12.67               & 10.99     \\ \hline
February  & 11.60              & 11.99                & 9.33             & 15.02             & 11.79      & 7.14         & 14.31               & 12.48     \\ \hline
March     & 11.52              & 11.21                & 9.10             & 14.32             & 12.40      & 6.82         & 13.49               & 11.96     \\ \hline
April     & 11.74              & 11.49                & 9.16             & 14.70             & 13.24      & 6.98         & 14.09               & 12.55     \\ \hline
May       & 11.58              & 11.91                & 9.16             & 15.14             & 12.34      & 7.17         & 15.00               & 13.11     \\ \hline
June      & 10.97              & 12.01                & 9.05             & 15.14             & 13.19      & 7.30         & 15.56               & 13.27     \\ \hline
July      & 10.80              & 11.95                & 9.00             & 15.07             & 13.13      & 7.30         & 15.57               & 13.19     \\ \hline
August    & 10.79              & 11.97                & 9.01             & 15.13             & 13.71      & 7.32         & 15.65               & 13.25     \\ \hline
September & 10.67              & 11.86                & 9.02             & 14.99             & 12.40      & 7.32         & 15.53               & 13.11     \\ \hline
\end{tabular}
    
    \label{tab:langcomp}
\end{table*}
\subsection{Hate and Stigma}
% claws training results
We trained the proposed model for hate recognition on the CLAWS dataset.
The implementation was done in PyTorch 1.7, and the optimization was performed with the Adam algorithm for 50 epochs.
The results are shown in Table \ref{tab:claws_net} and indicate the effectiveness of this network in monitoring hatefulness in tweets, improving upon the classifiers proposed for this task.
Given the large size of the dataset, which included over $2.5$ million tweets, and the complexity of the proposed model, instead of following a K-fold validation protocol, as done in \cite{ziems2020racism} for much simpler architectures, we evaluated our results on a single large hold-out set to measure the performance.

Utilizing this trained network, we created per-tweet assessments and aggregated them to create spatio-temporal trajectories marked with events.
Figure \ref{fig:median_counterhate} shows the pattern of the daily average value of counter-hate probability in the two states of California and New York.
The counter-hate probability is the probability of a tweet denouncing a hateful opinion.
An example of a tweet classified as counter-hate is shown in Figure \ref{fig:my_label}.
The inference on this dataset is made by employing the described hate assessment model which we have trained on CLAWS task.
The plots in Figure \ref{fig:median_counterhate} indicate an increase in the average daily counter-hate probability score per tweet during the span of $30$ days before to $30$ days after the start of Black Lives Matter protests. 

\begin{table}[]
\centering
\caption{Macro-averaged Performance Metrics for our model on the three CLAWS task and the AUROC score reported in \cite{ziems2020racism} for BERT-based classification.}
\begin{tabular}{|c|c|}
\hline
\multicolumn{2}{|c|}{\textbf{Our Model}}                       \\ \hline
Precision                          & 91.92                     \\ \hline
Recall                             & 94.83                     \\ \hline
F1-Score                           & 93.32                     \\ \hline
AUROC                            & 99.82                     \\ \hline
\multicolumn{2}{|c|}{\textbf{CLAWS baseline (AUROC)}} \\ \hline
Hate                          & 86.4                  \\ \hline
Counter-hate                  & 83.3                  \\ \hline
Neutral                       & 82.3                  \\ \hline
\end{tabular}
\label{tab:claws_net}
\end{table}

\begin{figure}[H]
    \centering
    \begin{lstlisting}[basicstyle=\ttfamily \footnotesize]
    
tweet (raw): We need to put negative comments about coronavirus aside you must realize it is not about who faults. We really don't know who faults it is. Stop attacked Asians because you can not tell difference between one over another. We need to find cure and help each other. Stop racist

hate distribution: (neutral: 0.0002, hate: 0.0001, counter-hate: 0.9997, other: 0.0001)
    
    \end{lstlisting}
    \caption{Example tweet classified as counter-hate}
    \label{fig:my_label}
\end{figure}

\subsection{Spatio-temporal Trajectories}
To better understand the dynamics of the conversations surrounding COVID-19, we aggregate our computed inferences on tweets per state and per day. This led to a series of spatio-temporal trajectories with a daily resolution. Our results indicate that for various groups of regions, these data can help further illustrate the patterns and directions of conversations on the COVID-19 during the course of the outbreak.

We focused on geo-tagged tweets, meaning that the metadata information regarding the US state from which every tweet was authored is present in the dataset.
We normalized our metrics by the overall output for each region so that we could reliably compare regions since larger or more populated states such as California have a larger volume of tweets than a smaller or less populated state such as Maine.
Our results indicate that following the trends around major news events can be very informative.
Figures similar to Figure \ref{fig:median_counterhate} and \ref{fig:langcomplexity_nynj} can be constructed using our platform, providing public health experts with an overview of critical pandemic-related temporal patterns.
More results on spatio-temporal trajectories are available in the appendix.\footnote{The full table of spatio-temporal features will be available in our codebase.}

In addition, we can extract informative auxiliary information by computing descriptive statistics such as count on our aggregate measurements across time and space. For example, Figure \ref{fig:symptom_count} shows the fluctuations in the average number of phrases likely to be describing symptoms present in the tweets across the US.

The fact that the curve flattens and plateaus after the first few months is consistent with the intuitive idea that the speculations about the specific characteristics of COVID-19 were more intensive when it was initially discovered.
As expected, the peaks fall in the first few months as people focus on the common symptoms expected to be experienced due to COVID-19 during the early stages of the pandemic.

\subsection{Limitations}

Given the sensitive nature of research in this area, it is critical to have an in-depth discussion of each approach's limitations:

\begin{itemize}
\item{
There are potential gaps in the data collected from the Twitter API due to privacy measures from Twitter, network errors, and/or API errors.
The same goes for the Reddit data.
}
\item{It is possible for the aforementioned RTN-based methodology of searching for COVID-19 entities to incorrectly tag phrases due to structural ambiguity, for example: "{\it covid-19 is really damaging. it is genuinely \textbf{hard to see}}" leads to the substring "{\it hard to see}" being returned as an instance of impact on activity. While our approach tremendously reduces the need for human supervision, a minimal qualitative assessment is still required.}
\item{
Designing LDA-based topic modeling frameworks, as well-recognized as they are, has known limitations.
It should be noted that the quantitative measures of the topic fitness, such as perplexity, are not always in accordance with human interpretation \cite{chang2009reading}.
}
\item{
We utilized different measures of language complexity as a {\it relative} assessment of readability. The absolute values for these methodologies and interpretations rely heavily on the nature of the text they process. For example, it is likely for a tweet not to have proper punctuation. 
While these limitations may lead to translations in the baseline, the results of relative assessments are consistent.
}

\item{
In analyzing and comparing the spatio-temporal curves, one should be mindful of the difference in the number of samples used in computing each aggregate datapoint, even when normalizing the data. 
Though we normalize the aggregates by the volume of tweets per region and time-point, a state with a much higher volume of tweets is more likely to capture a representative set of tweets than a state with a much lower volume of tweets.
For instance, the number of tweets made in California in the COVISION dataset is $15$ times more than the number of tweets made in Kentucky.
In comparing different regions, therefore, it is better to choose places for which the aggregate values are computed with a similar number of samples. 
This is why, in presenting some of our critical results, we focused on states with higher populations, such as New York and California, from which a larger number of tweets exist in our corpus.
}
\end{itemize}

\section{Conclusion}
% edited by shayan
This study methodically gathered a comprehensive dataset on social media conversations around the COVID-19 pandemic.
We focused on Reddit for acquiring comments that share experiences with COVID-19.
Additionally, we collected  geo-tagged tweets from across the United States that cover a broad time period starting at the beginning of the pandemic.
We proposed efficient search and analysis methodologies based on text structure and machine learning to better understand and assess the spatio-temporal trajectories of social media conversations on the novel coronavirus pandemic and empirically validated the performance of the proposed approach.
The framework of analysis we propose offers an automated pipeline that produces informative insights that requires minimal manual input.
This work demonstrates the effectiveness of analyzing social media data in order to extract critical information and knowledge with respect to public health interests during the COVID-19 pandemic.

% \begin{acks}
% \end{acks}

\bibliographystyle{./bibliography/IEEEtran}
\bibliography{./bibliography/refs.bib}

\clearpage
\appendix
\subsection{Data Acquisition}
The search terms used in our data acquisition scheme are shown in Table \ref{tab:searchterms}.

\begin{table}[H]
\caption{Query terms used in our data acquisition scheme}
\label{tab:searchterms}
\begin{tabular}{lll}
\multicolumn{3}{c}{\textbf{Query Terms}}                         \\ \hline
corona.*virus         & kungfuflu            & coronatoiletpaper \\
ncov                  & chingchongprague     & socialdistancing  \\
covid†                & commiecough          & social distancing \\
sars.*cov             & wuflu                & herdimmunity      \\
coronaalert           & miss rona            & quarentine        \\
corona.*outbreak      & ms. rona             & quarantine        \\
kung.*flu             & coronaviruschallenge & quarantinelife    \\
wuhan                 & coro                 & quarentinelife    \\
coronavirusapocalypse & coro coro            & coronacurfew      \\
pandemic              & cororo               & chinadisease      \\
epidemic              & miss coco v          & wuhanpneumonia    \\
quarantine            & la rona              & pandumbic         \\
rona                  & miss corona v        & chinacorona       \\
commie cough          & corov                & washhands         \\
wu-hanic plague       & corov19              & wash your hands   \\
mad-cau disease       & corov-19             & stayhome          \\
chinese virus         & corovid19            & stay at home      \\
chinesevirus          & corovid-19           & 6 ft              \\
chingchongvirus       & cocov                & 6 feet            \\
kungflufighting       & coronuh              & \#rona           
\end{tabular}
\end{table}

\subsection{Spatio-temporal Trajectories around Events \label{apx:spatiotemporal}}
In this section of the appendix, several other potential indicators of significant relationships are shown. Figure \ref{fig:median_hate_appendix1} indicates an increase in the average probability of a tweet in California and New York to exhibit COVID-related hateful sentiment. The increase could be associated with the considerable events in that time-span, such as the worsening of COVID-19 across US. A similar increase is shown in the onset of the pandemic as well, as visible in Figure \ref{fig:claws_hate_app1}.
\begin{figure*}[h]
    \centering
    \includegraphics[width=1\textwidth]{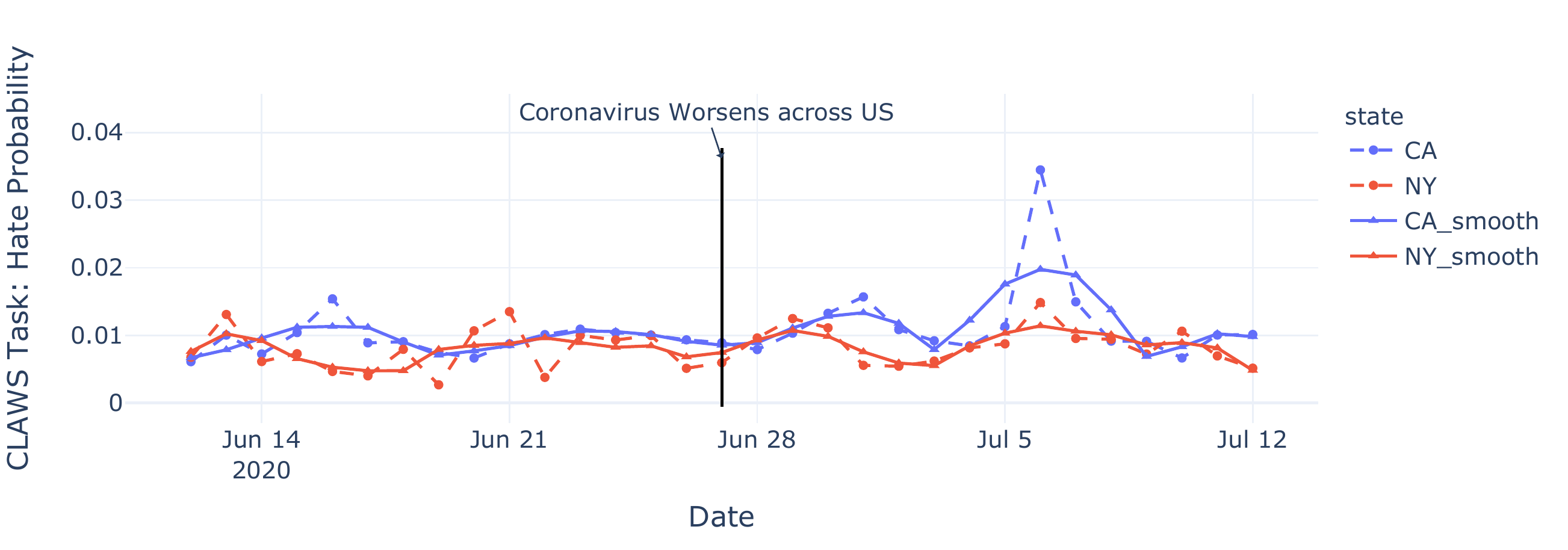}
    \caption{Daily averages of CLAWS Task: Hate probability around event: COVID-19 pandemic worsens across the US - Half Window: 15 days}
    \label{fig:median_hate_appendix1}
\end{figure*}
\begin{figure*}[h]
    \centering
    \includegraphics[width=1\textwidth]{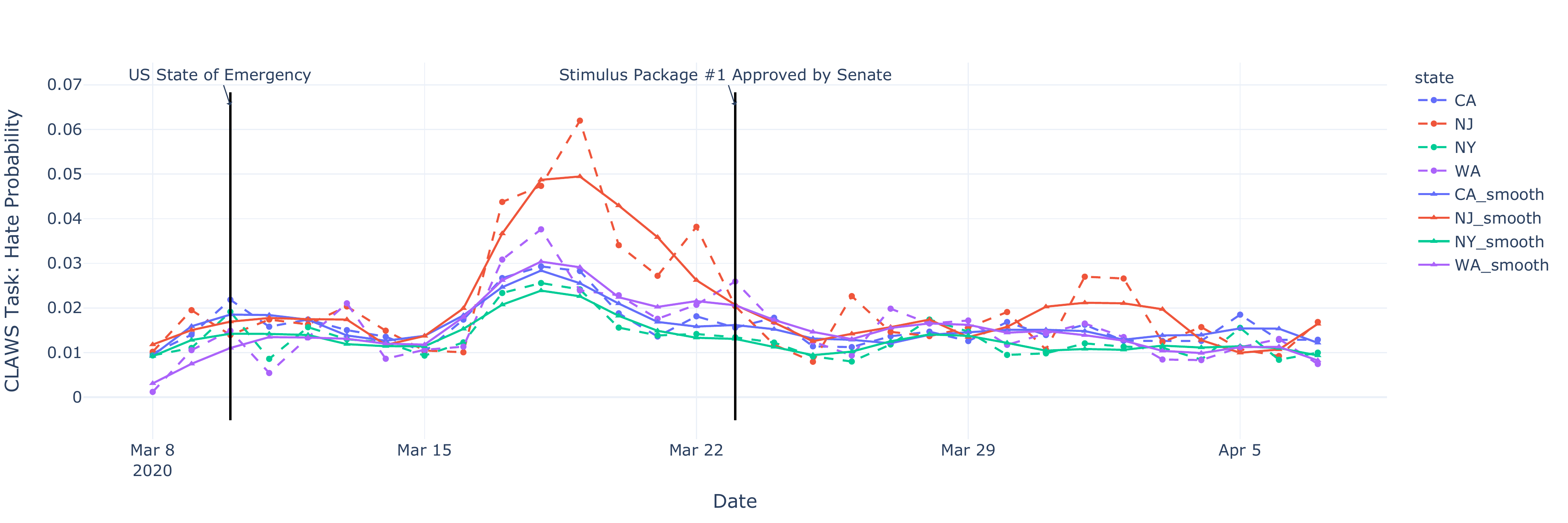}
    \caption{Daily average of CLAWS Task: Hate probability in the onset of the pandemic}
    \label{fig:claws_hate_app1}
\end{figure*}
\subsection{Language Complexity Metrics}\label{apx:langcomplexity}
\begin{figure*}[h]
    \centering
    \includegraphics[width=1\textwidth]{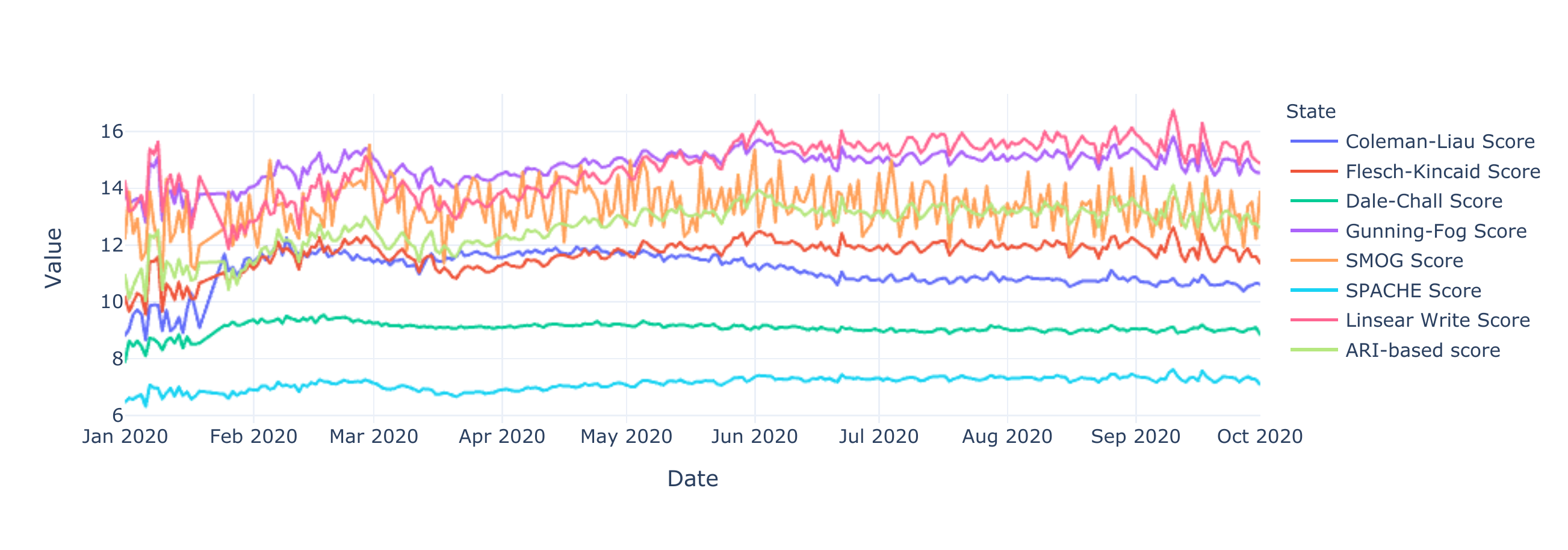}
    \caption{Daily Average of Measures of Readability Through Time}
    \label{fig:langcomplexity}
\end{figure*}

\subsubsection{Linsear Write Metric}
Developed for the United States Air Force, the linsear write readability metric was used to provide statistics on the readability of technical manuals. This test depends mainly on sentence length, and word syllables \cite{newsairforce}.
This algorithm's main intuition is that shorter sentences and shorter words are believed to be easier to read \cite{o1966gobbledygook}.

The algorithm for a $100$ word sample is as follows:
\begin{algorithm}
\caption{Linsear Write Readability Metric}
\label{CHalgorithm}
\begin{algorithmic}[1]
\Procedure{Linsear\textendash Write}{Text}
\State output $\leftarrow$ 0
\State sentCount $\leftarrow$ 0
\For{each sentence $s$ $\in$ Text}
\State sentCount $\leftarrow$ sentCount + 1
\For{each word $\omega \in s$}
\If{getSyllablesCount($\omega$) $<= 2$}
\State output $\leftarrow$ output $ + 1$
\Else
\State output $\leftarrow$ output $ + 3$
\EndIf
\EndFor
\EndFor

\If{output $> 20$}
\State output $\leftarrow$ output $ / 2$
\Else
\State output $\leftarrow$ output $ / 2 - 1$
\EndIf

\State \Return output

\EndProcedure
\end{algorithmic}
\end{algorithm}

\subsubsection{Dale-Chall Readability Test}
This numeric readability metric provides a measure on the difficulty and comprehensibility of the given text \cite{dale1948formula}.
Given a list of difficult words, the Dale-Chall formula is as follows:
\begin{multline}
\text{DC} = 0.0496 \times \cfrac{\text{Number of Words}}{\text{Number of Sentences}} \\ + 
0.1579 \times \cfrac{\text{Number of Difficult Words}}{\text{Number of Words}}
\end{multline}

\subsubsection{Coleman–Liau Readability Test}
The Coleman-Liau test is a language complexity metric that approximates the US grade level necessary to comprehend the text.
The formula for this test is as follows:

\begin{multline}
    \text{CLI} = 5.88 \times \cfrac{\text{Number of Letters}}{\text{Number of Words}} \\ 
    - 29.6 \times \cfrac{\text{Number of Sentences}}{\text{Number of Words}} - 15.8
\end{multline}

\subsubsection{SMOG Grade}
The Simple Measure of Gobbledygook (SMOG) Grade was developed as a readability test that approximates the years of education necessary to comprehend an input text. This readability test is widely used in the health domain as well \cite{ley1996use,hedman2008using}.
The formula for this test is as follows:

\begin{multline}
\text{SMOG} = 3.1291 \\ + 1.0430 \times \\ \sqrt{
\text{Number of Polysyllables} \times \cfrac{30}{\text{Number of Sentences}}
}
\end{multline}

\subsubsection{Gunning Fog Index}
The Gunning Fog Index is a readbility test to measure the number of years of formal education needed to understand the input text on the first read. This formula defines complex words as words with more than three syllables, and the output score is computed as follows:

\begin{multline}
    \text{GFog} = 0.4 \times (
    \cfrac{\text{Number of Words}}{\text{Number of Sentences}} \\ + 100 \times \cfrac{\text{Number of Complex Words}}{\text{Number of Words}}
    )
\end{multline}

\subsubsection{ Flesch-Kincaid Readability Tests}
Several instances of the aforementioned readability tests relied on the ratio of syllables to describe the text complexity. The Flesch-Kincaid readability tests also focus on these \cite{kincaid1975derivation}, with the reading ease test computed as follows:

\begin{multline}
\label{formula2}
    \text{FRES} = 206.835 - 1.015 \times \cfrac{\text{Number of Words}}{\text{Number of Sentences}} \\ - 84.6 \times \cfrac{\text{Number of Syllables}}{\text{Number of Words}}
\end{multline}

The higher the score, the easier the input text is expected to be in terms of readability. The modified version of the Equation \ref{formula2} is used to estimate the US grade level needed to understand this text, and it is shown in Equation \ref{formula3}.

\begin{equation}
    \label{formula3}
    0.39 \times \cfrac{\text{Number of Words}}{\text{Number of Sentences}} + 11.8 \times \cfrac{\text{Number of Syllables}}{\text{Number of Words}} - 15.59
\end{equation}

\subsubsection{Automated Readability Index}
The Automated Readability Index (ARI) is another test to approximate the grade level required to understand a text \cite{smith1967automated}. Similar to the aforementioned metrics, it is computed as follows:

\begin{multline}
    \text{ARI} = 4.71 \times \cfrac{\text{Number of Characters}}{\text{Number of Words}} \\ + 0.5 \times \cfrac{\text{Number of Words}}{\text{Number of Sentences}} - 21.43
\end{multline}
After rounding up the computed index, the value can be used to estimate the grade level. Higher values indicate a more difficult input text.

\subsubsection{SPACHE Readability Formula}
This measure of language complexity, which in nature is very similar to Dale-Chall algorithm, works by checking the words against a list of well-known words, and any word not appearing on that list is counted once and will be employed in computing the output. This measure is basically prepared for assessing primary texts \cite{10.2307/998915}.

\subsection{Topics \label{apx:lda}}
In addition to the per-topic distributions shown in Figure \ref{fig:twitter_lda_topics_barplots}, the word-clouds for each topic is also shown in Figure \ref{fig:twitter_lda_topics_wordclouds}.

\begin{figure*}
    \centering
\includegraphics[width=0.4\textwidth]{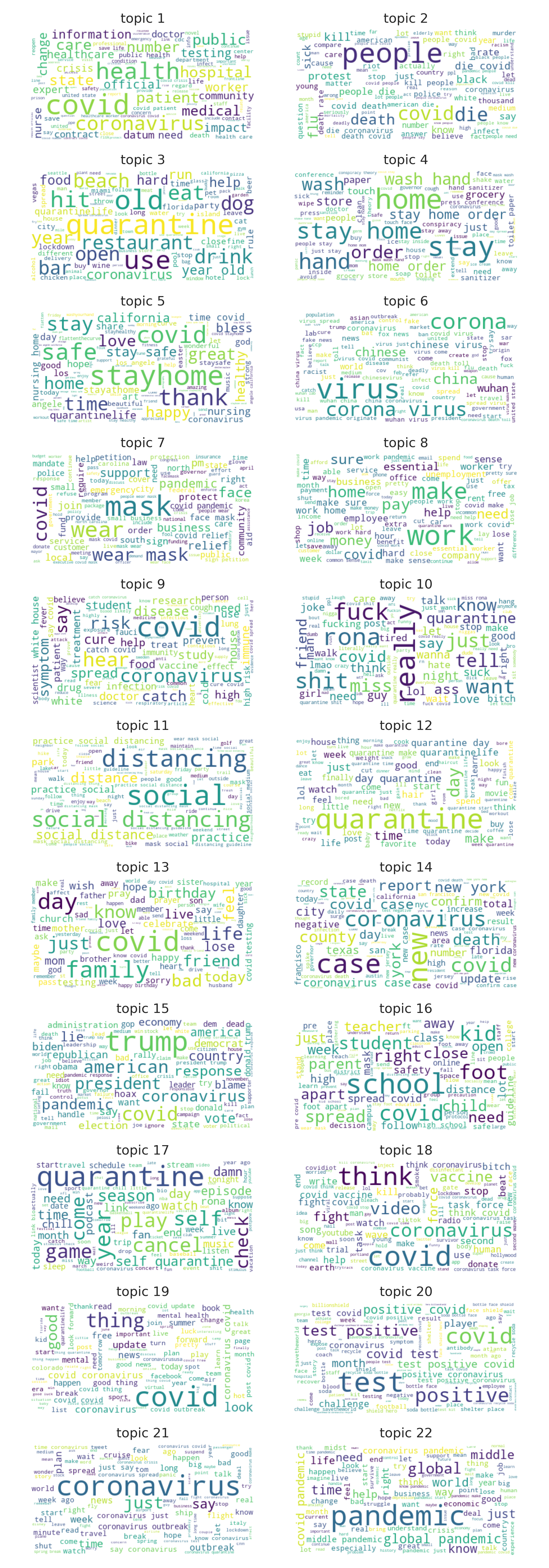}
    \caption{Wordclouds of the main topics of conversation in the first 9 months of 2020 across the US}
    \label{fig:twitter_lda_topics_wordclouds}
\end{figure*}

\end{document}